\documentclass[reprint,showpacs,nopreprintnumbers,amsmath,amssymb,superscriptaddress,aps,pra]{revtex4-1} 
\usepackage{color}
\usepackage{graphicx}
\usepackage{dcolumn}
\usepackage{bm}
\usepackage{gensymb} 
\usepackage{ifthen}

\DeclareMathOperator{\erf}{erf}

\begin{document}
\newcounter{wherearethereferences}
\setcounter{wherearethereferences}{3} 

\hyphenation{acce-le-ra-tion tomo-graphy prio-ri maxi-mum accor-dingly}
\newcommand{\etal}{\textit{et al.\ }}
\newcommand{\mat}[1]{{\mathbi{#1}}}
\newcommand{\vect}[1]{\mathbf{#1}} 
\newcommand{\vectgreek}[1]{\boldsymbol{#1}} 
\newcommand{\Vector}[2]{\genfrac{(}{)}{0pt}{}{#1}{#2}}
\newcommand{\tVector}[2]{\genfrac{(}{)}{0pt}{1}{#1}{#2}} 
\newcommand{\iec}{i.e., }
\newcommand{\ie}{\iec}
\newcommand{\egc}{e.g., }
\newcommand{\eg}{\egc}
\newcommand{\di}{d} 
\newcommand{\e}{e} 
\newcommand{\I}{i} 
\newcommand{\eq}{Eq.~} 
\newcommand{\eqs}{Eqs.~} 
\newcommand{\fig}{Fig.~} 
\newcommand{\figs}{Figs.~} 

\newcommand{\thisarticle}{this article~}  

\title{Two-center interference and ellipticity in high-order harmonic generation from H$_2^+$}

%

\author{Elmar V. \surname{van der Zwan}}
\email{ezwan@itp.uni-hannover.de}
\affiliation{Institut f\"ur Theoretische Physik and Centre for Quantum Engineering and Space-Time Research (QUEST), Leibniz Universit\"at Hannover, Appelstra\ss e 2, D-30167 Hannover, Germany}
\affiliation{Institut f\"ur Physik, Universit\"at Kassel, Heinrich-Plett-Stra\ss e 40, D-31432 Kassel, Germany}
\author{Manfred Lein}
\affiliation{Institut f\"ur Theoretische Physik and Centre for Quantum Engineering and Space-Time Research (QUEST), Leibniz Universit\"at Hannover, Appelstra\ss e 2, D-30167 Hannover, Germany}

\date{\today}

\begin{abstract}
We present a theoretical investigation into the two-center interference in aligned H$_2^+$. The influence of the laser field on the recombination step is investigated by comparing laser-induced harmonic generation with harmonic generation from field-free collisions of Gaussian wave packets with the core. We find that for different Gaussian wave packets colliding with the molecule, the interference minimum occurs at the same alignment angle. The same result is obtained for the laser-induced spectrum when only a single electronic trajectory per harmonic contributes. When multiple electronic trajectories contribute, we find an effect on the minimum position because the interference between short and long trajectories is alignment-dependent. The two-center interference and the influence of the Coulombic potential are clearly seen not only in the harmonic intensity and phase but also in the polarization direction and ellipticity. We observe significant ellipticity of the emitted radiation around the two-center interference minimum.
\end{abstract}


\maketitle

\section{Introduction}

When a gas of atoms or molecules is subjected to a strong laser field, high-harmonic generation (HHG) takes place \cite{Ferray88}. This process converts many of the laser photons into a single high-frequency photon in the extreme ultraviolet (XUV) or soft x-ray regime. The generation process can be understood in terms of the three-step model \cite{Corkum93}. First, the electron tunnels out under the influence of the laser field, and then it propagates freely in the laser field and can be driven back to the nucleus, where it can finally recombine. HHG has been used to generate coherent XUV radiation \cite{Hentschel01, Paul01}. Concerning HHG from molecules, there has been a lot of attention of the community recently toward the relationship between the emitted radiation and molecular characteristics. One can for instance determine the internuclear distance in the diatomic molecules H$_2^+$, H$_2$ and to a lesser extent CO$_2$ and O$_2$ from a destructive interference minimum in the harmonic spectrum \cite{Lein02_2, Lein02, Le09_3} or reconstruct molecular orbitals from the emitted radiation \cite{Itatani04, Haessler10}. The correct explanation of the minimum in CO$_2$, however, appears to require a multiorbital treatment \cite{Smirnova09}. Recent experiments on N$_2$ indicate multiorbital contributions as well \cite{McFarland08}. Preparing molecules to control the harmonic emission is also possible: ring current states were used theoretically to generate circularly polarized high harmonics \cite{Xie08}. 

In the three-step model \cite{Corkum93} or its quantum-mechanical formulation in terms of the strong-field approximation (SFA) \cite{Lewenstein94}, due to wave-packet spreading the on-axis continuum wave packet carries practically no signature of the ground-state wave function at the moment of recombination. Therefore, the molecular characteristics imprinted on the emitted spectra must come from the recombination step (or from multiorbital interference). In the existing practical molecular imaging techniques the influence of the laser field on the recombination step and often also the effect of the Coulomb potential on the propagation step are ignored. To study the effects of these approximations, we perform a numerical comparison between the harmonics emitted in a normal laser-induced HHG process and harmonics emitted when an artificially prepared wave packet collides with the molecular ion in the absence of any laser pulse \cite{Lein02_2}. Additionally, we compare the harmonics generated by both physical and artificial laser pulses to disentangle the influence of the different harmonic trajectories. We show that elliptically polarized HHG radiation from linearly polarized generating pulses occurs near the two-center interference. Elliptical polarization of harmonics from linearly polarized generating pulses was also found in recent experiments \cite{Zhou09}. 

\section{Method}

In \thisarticle we solve the time-dependent Schr\"odinger equation (TDSE) numerically in two dimensions for a molecular ion with a single electron. We consider two dimensions because in 2D many of the 3D characteristics of HHG are already present, such as the existence of directions perpendicular and parallel to the laser polarization axis. On the other hand, the TDSE can be solved very quickly, allowing the TDSE to be solved for many alignment angles of the molecule in the laser field. The TDSE is solved using the split-operator method \cite{Fleck76,Feit82}, and the ground-state wave function is found by imaginary-time propagation \cite{Kosloff86}. 

Atomic units are used throughout \thisarticle unless indicated otherwise. We will focus on 2D H$_2^+$ with fixed nuclei. The TDSE reads
\begin{subequations}
\begin{align}
i\frac{\partial\psi(\vect{r},t)}{\partial t}&=\hat{H}\psi(\vect{r},t),\\
\hat{H}&=\frac{\hat{\vect{p}}^2}{2}+ V(\vect{r}) + \vect{r}\cdot \vect{E}(t), \label{eq: H with laser} \\
V(\vect{r}) &= -\frac{1}{\sqrt{(\vect{r}-\frac{\vect{R}}{2})^2+a^2}}- \frac{1}{\sqrt{(\vect{r}+\frac{\vect{R}}{2})^2+a^2}},
\end{align}
\end{subequations}
where $\hat{\vect{p}}$ is the momentum operator, $\vect{r}=\tVector{x}{y}$, $V$ is a softcore potential, and $\vect{E}(t)$ is the time-dependent electric field of the laser pulse. The internuclear axis $\vect{R}$ makes an angle~$\theta$ with respect to the laser polarization axis~$x$. Using the softcore parameter $a^2=0.5$ and an internuclear distance of $R=2$, the ionization potential is $I_\mathrm{p}=30.2$ eV. The total harmonic emission spectrum including $x$- and $y$-polarization is calculated from the numerical solution of the TDSE as
\begin{subequations}
\begin{align}
S(\omega) &= \vert \vectgreek{\alpha}(\omega) \vert^2, \quad \vectgreek{\alpha}(\omega) = \int W(t) \langle \vectgreek{\alpha}(t) \rangle \e^{\I \omega t} \di t, \label{eq: TDSE harmonics alpha}\\
\langle \vectgreek{\alpha}(t) \rangle &= \langle \psi(\vect{r},t) \vert \vectgreek{\nabla} V(\vect{r}) + \vect{E}(t) \vert \psi(\vect{r},t) \rangle,    
\end{align}
\label{eq: TDSE harmonics}
\end{subequations}
where $\langle \vectgreek{\alpha}(t) \rangle$ is the dipole acceleration and $S(\omega)$ is proportional to the intensity of the emitted radiation at frequency $\omega$. Here $W(t)$ is a standardly used temporal window that prevents high-frequency artifacts at the boundaries of the integration.

For H$_2^+$, one observes a minimum in the spectrum of emitted radiation polarized in the $x$-direction because of interference between the two centers of the molecule \cite{Lein02_2}. Using the plane-wave approximation for the returning electron, the (first and usually only observable) minimum occurs when the $x$-projection of the internuclear distance as seen by the returning wave packet is equal to half the de Broglie wavelength. Therefore the minimum will shift toward higher harmonics with higher angles between the laser polarization and molecular axis. Since it is a structural minimum that depends only on the geometry of the bound state, one expects to see no shifts in the location of the minimum when different laser pulses are used or when instead Gaussian wave packets are used to generate harmonics in a laser-field-free electron-ion collision \cite{Lein02_2}. 

For the wave-packet simulation without laser field, the center of the potential is placed in the middle of the grid at $(x,y)=(0,0)$, and the Gaussian wave packet is introduced with its center at position $(x_0=L_x/4,0)$, with the grid size denoted as $L_x \times L_y$. The initial wave packet $\psi(\vect{r})$ is given by the superposition
\begin{subequations}
\begin{align}
\psi(\vect{r}) &= \psi_0(\vect{r}) + \psi_\mathrm{G}(\vect{r}),\\
\psi_\mathrm{G}(\vect{r}) &= \sqrt{C}\sqrt{\frac{c_x c_y}{\pi}}\e^{-\frac{1}{2}\left(c_x^2 \left(x-\frac{L_x}{4}\right)^2+c_y^2 y^2\right)+\I k_0 x}, 
\end{align}
\end{subequations}
where $\psi_0(\vect{r})$ is the ground-state wave function, and $c_x$, $c_y$ quantify the momentum spread of the Gaussian wave packet in the $x$- and $y$-directions. The wave packet moves with a central momentum $k_0<0$ toward the molecular core. The norm of the Gaussian wave packet $C$ should be set small to mimic the situation of HHG at the typically used intensities. We use $C= 10^{-6}$. The momentum-spread parameter in the $y$-direction is chosen as 
\begin{equation}
c_y = r_k |k_0|,
\end{equation}
where a tuning parameter $r_k$ is used to study the effects of the different types of Gaussian wave packets and can be set to simulate the character of the continuum wave packet as generated by a laser pulse. The momentum-spread parameter in the $x$-direction $c_x$ is set relatively large to allow for many harmonics to be probed by one wave packet. The propagation time is chosen such that a classical particle with momentum $k_0$ moves from $(L_x/4,0)$ to $(-L_x/4,0)$ during the propagation. As a result, the strongest emission is expected at the middle of the propagation, such that little distortion is introduced when using a window function in the temporal Fourier transform for obtaining the power spectrum. As an example, for $r_k=0.01$ and $k_0=-1.78$, the grid dimensions are $L_x=383$ a.u.\ and $L_y=1006$ a.u. Here we use $2304\times 6144$ spatial grid points and 2000 time steps. The propagation time equals the optical period of a laser field with a 780-nm wavelength.   

For the simulation of the laser-induced HHG process we use a laser pulse linearly polarized in the $x$-direction. The initial state is set to the ground state $\psi_0(\vect{r})$. The time-dependent wave function is propagated for the laser pulse duration and two additional cycles after the end of the laser pulse to minimize distortions from the dipole acceleration window and to allow the wave packets to return to the nucleus. For the laser-induced calculations we use a grid measuring $280 \times 84$ a.u.\ with $1536 \times 512$ grid points and 2000 time steps per optical cycle.

\section{Harmonic intensity}

\begin{figure}[tbp]
\includegraphics[width=0.8\columnwidth]{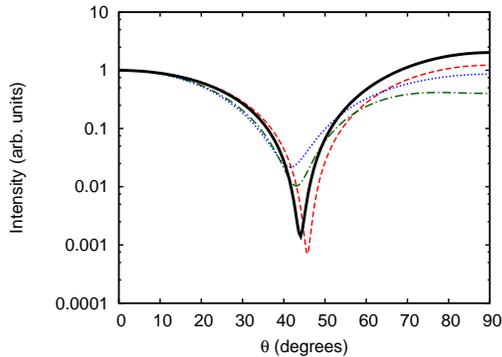}
\caption{(Color online) Intensity of harmonic 49 vs alignment angle $\theta$ for emission polarized along the $x$-direction. A Gaussian wave packet (black solid line) is compared to a three-cycle $\sin^2$-pulse (red dashed line), a five-cycle $\sin^2$-pulse (blue dotted line) and a fifteen-cycle trapezoidal pulse (green dot-dashed line).}
\label{fig: intensity1}
\end{figure}

The laser wavelength for the calculations of laser-induced HHG is 780 nm. For the Gaussian-wave-packet collisions, the same energy scale in units of harmonics of a 780 nm laser pulse is used. In \fig\ref{fig: intensity1} we show the alignment dependence of the emitted radiation for harmonic 49 polarized along the $x$-direction. We compare a Gaussian wave packet with $k_0=-1.78$ and $r_k=0.01$ to laser pulses with an intensity of $I=5\times 10^{14}$~W/cm$^2$ and different lengths. The laser pulses have a $\sin^2$ envelope of either three or five cycles, or a trapezoidal envelope of fifteen cycles length with five-cycle ramps. The carrier-envelope phase, \ie the phase between the carrier wave and the envelope, is $\tfrac{\pi}{4}$ for the shortest pulse and 0 for the five-cycle $\sin^2$-pulse. The trapezoidal pulse has a sinusoidal carrier wave. The intensity data are integrated over one harmonic order. The figure shows that the different laser pulses give rise to minima that are near, but not exactly at, the same position as the minimum from the Gaussian wave packet.

\begin{figure}[tbp]
\includegraphics[width=1.0\columnwidth]{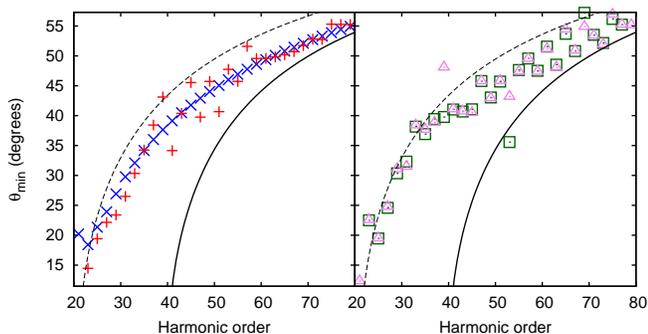}
\caption{(Color online) Location of the minimum in the alignment dependence of the intensity polarized along the $x$-direction. (Left) Blue crosses are for the Gaussian wave packet and red plusses for the three-cycle pulse of \fig\ref{fig: intensity1}. (Right) Green squares are for the fifteen-cycle pulse of \fig\ref{fig: intensity1} and violet triangles are for a ten-cycle trapezoidal pulse. The black solid line displays the two-center interference based on the SFA relation $k(\omega)=\sqrt{2(\omega-I_\mathrm{p})}$ and the black dashed line is the $I_{\mathrm{p}}$-corrected result based on the relation $k(\omega)=\sqrt{2\omega}$.}
\label{fig: intensity2}
\end{figure}

In \fig\ref{fig: intensity2} we plot the positions of the minima $\theta_{\mathrm{min}}$ in the alignment dependence versus harmonic order for different laser pulses and the Gaussian wave packet from \fig\ref{fig: intensity1}. Also indicated in the figure are the curves that are predicted for the two-center minimum ($R \cos\theta_{\mathrm{min}}=\pi/k$) \cite{Lein02} using either the energy-conserving relationship $k(\omega)=\sqrt{2(\omega-I_\mathrm{p})}$ from the Lewenstein model \cite{Lewenstein94}, or using the $I_\mathrm{p}$-corrected relationship $k(\omega)=\sqrt{2\omega}$ that has been adopted previously for molecular imaging \cite{Lein02_2, Itatani04}. The physical argument for the $I_\mathrm{p}$-correction is that, when describing the returning electron as a plane wave, one should take into account that at the moment of recombination its wave number is modified by the absorption of $I_{\textrm{p}}$ into the kinetic energy.
From \fig\ref{fig: intensity2} it is clear that the Gaussian wave packet gives rise to a very smooth shift of the minimum as a function of alignment angle $\theta$. The laser pulses produce minima that follow the same trend as the Gaussian wave packet, but are scattered around the general trend. The results for short and longer pulses are scattered differently, but not less or more. The differences between relatively long ten- and fifteen-cycle pulses pulses are small. This is expected, since both pulses are effectively almost cw-like. Even for the fifteen-cycle pulse the depletion of the ground state remains below 6\%. The results suggest that when using the plane-wave approximation for the returning electron in molecular imaging applications, a dispersion relationship in between the Lewenstein and $I_\mathrm{p}$-corrected relationships should be used \cite{Levesque07,Gonoskov08}.

\subsection{Effect of the propagation step}
\label{sec: intensity: propagation step}

\begin{figure}[tbp]
\includegraphics[width=1.0\columnwidth]{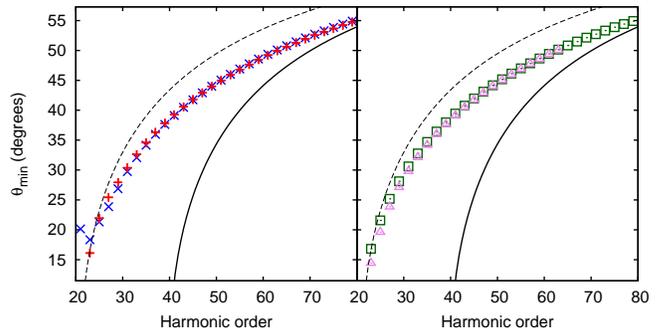}
\caption{(Color online) (Left) Same as \fig\ref{fig: intensity2} for a broad Gaussian wave packet starting far from the nucleus (blue crosses) and a broad Gaussian wave packet starting close to the nucleus (red plusses). (Right) A narrow Gaussian wave packet starting close to the nucleus (green squares) and a broad Gaussian wave packet starting far from the nucleus but with less energy (violet triangles). The blue crosses and red plusses have $r_k=0.03$ and $k_0=-1.78$, the violet triangles have the same $r_k$ but $k_0=-1.38$. The red plusses and green squares start at $x_0=20$~a.u., the blue crosses correspond to an initial position of $x_0=95.8$~a.u.\ and the green squares to $x_0=74.1$~a.u.}
\label{fig: intensity3}
\end{figure}

We compare the positions of interference minima for different types of Gaussian wave packets in \fig\ref{fig: intensity3}. We vary the momentum spread of the wave packet in the perpendicular direction, the central momentum of the wave packet and the position at which the wave packet starts. 
The striking and important observation is that all curves lie very close together. Apparently the position of the minimum is insensitive to the momentum distribution of the continuum wave packet. Only for the lowest harmonics we can observe some difference between the different kinds of continuum wave packets. Small differences appear there between wave packets starting far away and those starting close to the nucleus, due to the effect of the long-range binding potential. Our findings suggest that in terms of the three-step model, the propagation step has little effect on the observed position of the minimum and cannot account for the big fluctuations observed in \fig\ref{fig: intensity2}. 

\subsection{Effect of the recombination step}

In the three-step model \cite{Corkum93}, the laser field during the recombination step and possible interferences between different parts of the continuum wave packet are ignored. In reality, the electronic wave packet does not recombine under laser-field-free conditions, and different trajectories recombine at different times with different phases. To study the effect on the recombination process, we resort to a comparison of the minimum positions using artificial pulses. The pulses are four-cycle sinusoidal pulses with a constant envelope corresponding to an intensity of $I=5\times 10^{14}$ W/cm$^2$, \ie a section of a cw laser field. At $t=0$, the electric field is $E(0)=0$. Optionally, we employ either or both of two methods to influence the recombination step: (i) setting the dipole acceleration to 0 after some point in time during the propagation and (ii) turning off the laser field for the inner region near the nuclei after some point in time.

Every half laser cycle, both a short and long classical electronic trajectory contribute to every harmonic peak \cite{Lewenstein94}. The distinction between short and long trajectories is based on whether the electron spends shorter or longer than $0.65T$ in the continuum, where $T$ is the laser period. Setting the dipole acceleration to 0 beyond $t=T_\alpha$ using a temporal width $\Delta T_\alpha$ \cite{Smirnova09}, \eq\eqref{eq: TDSE harmonics alpha} becomes 
\begin{subequations}
\begin{align}
\vectgreek{\alpha}(\omega) &= \int_0^L W'(t) \langle \vectgreek{\alpha}(t) \rangle \e^{\I \omega t} \di t,\\
W'(t) &= W(t) S(t),\\
S(t) &= \begin{cases} 1 & \text{for $t \le t_1$}\\ \cos^2 \left(\frac{t-t_1}{t_2-t_1}\frac{\pi}{2}\right) &\text{for $t_1 < t < t_2$} \\ 0 &\text{for $t \ge t_2$,}\end{cases}
\end{align}
\label{eq: alpha filter}
\end{subequations}
where $L$ is the propagation length, $t_1=T_\alpha - \tfrac{\Delta T_\alpha}{2}$ and $t_2=T_\alpha + \tfrac{\Delta T_\alpha}{2}$. We use $\Delta T_\alpha=0.1T$. We set the dipole acceleration to 0 at either $T_\alpha=0.95T$, the return time of the most energetic trajectory, or at $T_\alpha=1.182T$, the time at which the return momentum of the first half-cycle's long trajectory matches that of the second half-cycle's short trajectory. Thus with $T_\alpha=0.95T$ we take into account only the short trajectories from the first half-cycle, and $T_\alpha=1.182T$ is the optimal point in time for selecting only a single pair of short and long trajectories.

Additionally, we optionally turn off the laser in the inner region at time $t=T_{\mathrm{l}}$. To prevent artifacts, the field is turned off gradually in both space and time. The laser interaction is completely turned off for $r=\sqrt{x^2+y^2} < 4$, undisturbed for $r > 6$, and we use a $\sin^2$-transition between these two extremes. In the time domain, we use a smoothened step function (convolution of a Gaussian with a step function) with a width of $0.1T$. In formula, the Hamiltonian in \eq\eqref{eq: H with laser} is replaced by
\begin{subequations}
\begin{align}
\hat{H}&=\frac{\hat{\vect{p}}^2}{2}+ V(x,y) +  Z(x,y,t) \cdot x E(t),\\
Z(x,y,t) &= F(x,y) + (1-F(x,y)) R(t),\\
F(x,y) &= \begin{cases} 0 &\text{for $r \le 4$} \\ \sin^2(\frac{\pi}{4}(r-4)) & \text{for $4 < r < 6$}\\ 1 & \text{for $r \ge 6$,} \end{cases} \\
R(t) &= \frac{1}{2}\left( 1-\erf\left(\left(t-T_{\mathrm{l}}\right)/\left(\sqrt{2} \cdot 0.1 T \right)\right)\right).
\end{align}
\end{subequations}
When $T_{\mathrm{l}}=0.5T$ is used, this special setup allows us to compare near-physical harmonics to those generated in an identical setup where only trajectories starting during the first half-cycle contribute and with the laser field completely turned off during all recombinations. This method could be easily extended to filter out either the short or long trajectories. 

\begin{figure}[tbp]
\includegraphics[width=1.0\columnwidth]{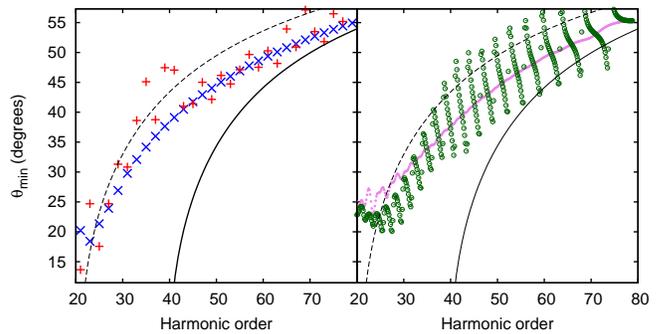}
\caption{(Color online) (Left) Same as \fig\ref{fig: intensity2} for a broad Gaussian wave packet starting far from the nucleus (blue crosses) and for harmonics generated from a sinusoidal laser pulse (red plusses). (Right) Harmonics from sinusoidal pulses with $T_\alpha=1.182T$ (green circles) and with $T_\alpha=0.95T$ (violet points). }
\label{fig: intensity4}
\end{figure}

\begin{figure}[tbp]
\includegraphics[width=0.8\columnwidth]{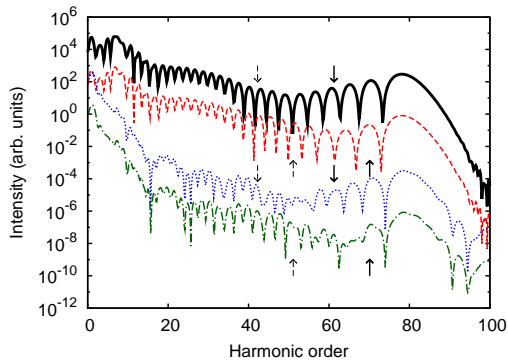}
\caption{(Color online) Harmonic intensity polarized in the $x$-direction for the green circles from \fig\ref{fig: intensity4} for $\theta=45\degree$ (black solid line) and $\theta=50\degree$ (red dashed line). Also for the violet points from \fig\ref{fig: intensity6} for $\theta=45\degree$ (blue dotted line) and $\theta=50\degree$ (green dot-dashed line). Solid and dashed arrows indicate spectral minimum positions predicted using $k(\omega)=\sqrt{2(\omega-I_\mathrm{p})}$ and $k(\omega)=\sqrt{2\omega}$, respectively.}
\label{fig: intensity5}
\end{figure}

In \fig\ref{fig: intensity4} we compare the scattering around the Gaussian-wave-packet results from the sinusoidal pulse (red plusses) with those from setting the dipole acceleration to 0 at $T_\alpha=1.182T$  (green circles) and from setting the dipole acceleration to 0 at $T_\alpha=0.95T$ (violet points). 
With $T_\alpha=1.182T$ the interference between the long and short trajectories leads to a strong, but regular oscillation of the laser-induced results around the Gaussian-wave-packet results. The same interference between the short and long trajectories can be seen in an associated harmonic spectrum as the top, black solid line in \fig\ref{fig: intensity5}. When every harmonic peak is caused by a single trajectory (violet points in \fig\ref{fig: intensity4}), the interference disappears completely and the result is almost as smooth as that from the Gaussian wave packet. Under normal circumstances, additional later returns from the same trajectories contribute to the spectrum. For a finite pulse length, different half-cycles also contribute differently because of the pulse envelope. Additionally, for a numerical calculation the dipole acceleration window $W(t)$ also changes the contributions between different half-cycles. All of these together then smoothen but irregularize the oscillation of the green circles in \fig\ref{fig: intensity4}, leading to the scattering of the laser-induced data points observed in \fig\ref{fig: intensity2}. The interference between the short and long trajectories can probably also explain the scattering of the two-center minimum as a function of intensity as found by Gonoskov and Ryabikin \cite{Gonoskov08}.

\begin{figure}[tbp]
\includegraphics[width=0.8\columnwidth]{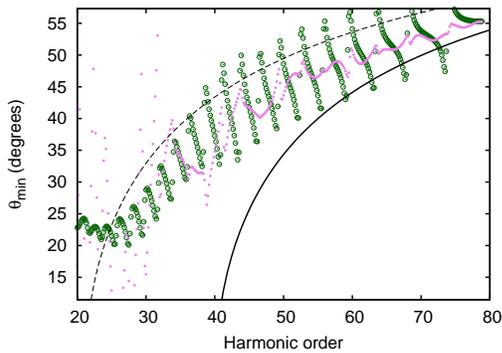}
\caption{(Color online) Same as \fig\ref{fig: intensity2} for harmonics from a sinusoidal laser pulse with $T_\alpha=1.182T$ (green circles) and additionally with $T_{\mathrm{l}}=0.5T$ (violet points). }
\label{fig: intensity6}
\end{figure}

The fact that the results for a single harmonic trajectory (violet points in \fig\ref{fig: intensity4}) lie so close to the Gaussian-wave-packet result, means that the laser field has no significant influence on the amplitude of the recombination matrix elements. This supports using HHG for molecular imaging \cite{Lein02, Lein02_2, Itatani04, Haessler10}, as in a typical experimental setup only short trajectories contribute to the harmonic spectrum. Interestingly, however, in \fig\ref{fig: intensity6} we show that turning off the laser field during the recombination does have a significant effect on the interference between the short and long trajectories. 
In the figure, the green circles are copied from \fig\ref{fig: intensity4}. Additionally, we show the case where there is only a single set of short and long trajectories with additionally the laser pulse turned off in the inner region at $T_{\mathrm{l}}=0.5T$ (violet points). The strong reduction in scattering amplitude for the violet points in \fig\ref{fig: intensity6} can be understood from \fig\ref{fig: intensity5}, where we plot the harmonic spectra for molecular alignment angles $45\degree$ and $50\degree$ for the case of the green circles (violet points) in \fig\ref{fig: intensity6} as the top (bottom) two curves. For the unmodified laser pulse we observe a significant shift of the trajectory interference positions in the harmonic spectrum when going from alignment at $45\degree$ to $50\degree$. Although at first sight the bottom two curves in \fig\ref{fig: intensity5} look more distinct from one another, a closer look reveals that the alignment dependence of the trajectory interference minima is actually a lot smaller with the laser field turned off in the inner region, as there is no shift visible. The strong scattering at low harmonics for the violet points in \fig\ref{fig: intensity6} is caused by the fact that the finite temporal widths of the filters $R(t)$ and $S(t)$ suppress the complete lower end of the spectrum.  

\section{Harmonic phase}

\begin{figure}[tbp]
\includegraphics[width=0.8\columnwidth]{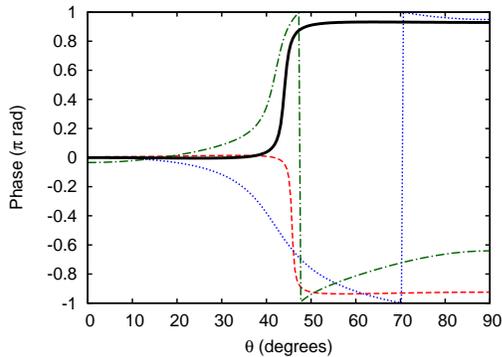}
\caption{(Color online) Phase of the harmonic emission polarized along $x$ for harmonic 49 generated by a Gaussian wave packet (black solid line) and generated by a three-cycle $\sin^2$-pulse (red dashed line), a five-cycle $\sin^2$-pulse (blue dotted line), and a fifteen-cycle trapezoidal pulse (green dot-dashed line).}
\label{fig: harmonicphase}
\end{figure}

The two-center minimum in the harmonic spectrum is accompanied by a phase jump in the harmonic phase. Using the plane-wave approximation, this should be a sharp $\pi$-phase jump \cite{Lein02}. However, in experiments a smaller and smoother phase jump is observed \cite{Boutu08}. Such deviations can be attributed to nonclassical momenta \cite{Chirila09} and to effects of the Coulombic potential \cite{Ciappina07}. Similarly, a phase jump is observable when one considers a fixed harmonic as a function of $\theta$. The phase of harmonic 49 for emission polarized along $x$ is shown in \fig\ref{fig: harmonicphase}. The same set of laser pulses and Gaussian wave packet is used as in \fig\ref{fig: intensity1}. The curves have been shifted such that for $\theta=0$ the phase is 0.  
The figure shows that both the Gaussian wave packet and the extremely short three-cycle laser pulse give rise to a mostly constant phase as a function of $\theta$ with a phase jump slightly smaller than $\pi$ at the location of the minimum. The longer pulses have a more smeared-out phase jump. In the neighborhood of the minimum their jump is a lot smaller than $\pi$ but over the complete $\theta$-range the jump seems to be bigger than $\pi$. This behavior for the longer pulses is probably an effect of more, and longer, trajectories contributing to the harmonics. Different trajectories are associated with different Coulomb corrections and therefore the harmonic phase becomes smeared out. This is in accordance with the shallower intensity minima in \fig\ref{fig: intensity1} for the longer pulses. 

\begin{figure}[tbp]
\includegraphics[width=0.8\columnwidth]{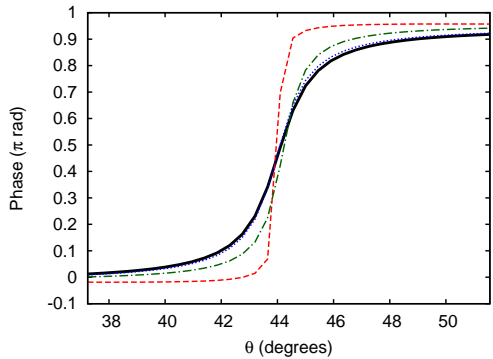}
\caption{(Color online) Phase of the harmonic emission polarized along $x$ for harmonic 49 generated by a broad Gaussian wave packet starting far from the nucleus (solid black line), a broad Gaussian wave packet starting close to the nucleus (red dashed line), a narrow Gaussian wave packet starting close to the nucleus (green dot-dashed line), and a broad Gaussian wave packet starting far from the nucleus but with less energy (blue dotted line).}
\label{fig: harmonicphase2}
\end{figure}

It is interesting to investigate the phase jump for the different Gaussian wave packets of \fig\ref{fig: intensity3}. This is plotted in \fig\ref{fig: harmonicphase2} for a smaller range of $\theta$ for clarity.
The broad Gaussian wave packet starting closing to the nucleus (red dashed line) starts out with small perpendicular momentum components. Because of the short propagation time before the interaction with the core, both Coulomb effects on the momentum distribution and perpendicular momentum components will be relatively small in this scenario. As a result we observe a sharp almost-$\pi$-phase jump. To some extent, the difference with the narrow Gaussian wave packet (green dot-dashed line) is that the latter experiences roughly the same Coulomb effects, but starts out with much larger perpendicular momentum components. This leads to a much smoother phase jump. Independent of the central momentum of the wave packet, a broad wave packet starting far away from the nucleus (black solid and blue dotted lines) experiences a smoother phase jump because of Coulomb effects and the associated increased nonparallel momentum components.

\section{Harmonic polarization}

High harmonic radiation is coherent with well-defined polarization \cite{Zhou09}. We can write the emitted radiation field $\vect{E}_\mathrm{em}(t)$ as
\begin{equation}
\vect{E}_\mathrm{em}(t)=\int_{-\infty }^{\infty}\frac{c(\omega)}{\sqrt{1+(\epsilon(\omega ))^2}}\left(\vect{e}_\mathrm{p}(\omega)  +\I \epsilon(\omega ) \vect{e}_\mathrm{o}(\omega)\right)e^{-\I \omega  t} \frac{d\omega}{2\pi},
\label{eq: generalpulse}
\end{equation}  
where $c$ is a complex function and $\epsilon$ is the ellipticity as a function of frequency $\omega$. Here $\epsilon$ can be either positive or negative, $\epsilon=0$ represents linearly polarized light, and $\epsilon=\pm 1$ represent positively and negatively circularly polarized light, depending on the choice of propagation direction. Furthermore, $\vect{e}_\mathrm{p}$ is the unit vector in the main polarization direction, and $\vect{e}_\mathrm{o}$ is the unit vector in the perpendicular direction. The angle between the main polarization direction of the emitted radiation and the polarization axis of the laser pulse is $\phi$. Similarly to the angle $\theta$ between the molecular axis and the laser polarization axis, a counterclockwise rotation corresponds to a positive angle. However, we limit the definition of $\phi$ to the range $[0,\pi)$, whereas the complex number $c$ covers the full complex domain. At a given $\omega$, all possible types of coherent plane-wave radiation can be uniquely described by the parameters $(\phi, \epsilon, c)$. 

Numerically the acceleration of the dipole moment is identified with the far-field harmonic field (up to an overall phase). Therefore the Fourier transformed dipole acceleration is given by
\begin{equation}
\vectgreek{\alpha}(\omega)=\frac{c(\omega)}{\sqrt{1+(\epsilon(\omega))^2}}\left(\vect{e}_\mathrm{p}(\omega)  +\I \epsilon(\omega) \vect{e}_\mathrm{o}(\omega)\right).
\label{eq: a(w) polarization}
\end{equation} 
For an experimentalist the most practical way to measure the polarization angle and ellipticity of the emitted radiation is to pass the harmonics through a polarization filter and measuring the emitted intensity for many polarization directions. The direction of greatest emission corresponds to the main polarization direction, and the emission in the orthogonal direction is a measure for the ellipticity of the emitted radiation \cite{Zhou09}. Only under considerable effort for the experimentalist, the harmonic phase can be measured  interferometrically as a function of alignment angle \cite{Smirnova09, McFarland09} or as a function of harmonic order \cite{Mairesse03}. Theoretically one has easy access to the harmonic phase. Using the phase information, \eq\eqref{eq: a(w) polarization} can be inverted as described in the following to deduce the polarization parameters $\phi$, $\epsilon$ and the complex number $c$ from the complex-valued $\alpha_x$ and $\alpha_y$. As a measure of the ellipticity we define a cross term $\sigma$ as (we omit the $\omega$-dependence for simplicity)
\begin{equation}
\sigma \equiv\vert \alpha_x \vert \vert \alpha_y \vert \sin\delta =\frac{\vert \vectgreek{\alpha} \vert^2 \epsilon}{1+\epsilon^2},
\label{eq: sigmadefinition}
\end{equation}
where \mbox{$\delta=\arg(\alpha_y)-\arg(\alpha_x)$} and we used that $\vert c \vert^2 = \vert \vectgreek{\alpha} \vert^2$. The following equalities can be straightforwardly derived for the polarization parameters in terms of $\vectgreek{\alpha}$ and $\sigma$,
\begin{subequations}
\begin{align}
\epsilon&=\frac{1-\sqrt{1-4\left(\frac{\sigma}{\vert\vectgreek{\alpha}\vert^2}\right)^2}}{2\frac{\sigma}{\vert\vectgreek{\alpha}\vert^2}},\\
\tan\phi &= \frac{\alpha_y - \I \: \epsilon \: \alpha_x}{\alpha_x + \I \: \epsilon \: \alpha_y},\\
c&=\begin{cases}
\frac{\sqrt{1+\epsilon^2}}{\cos\phi -\I \: \epsilon \sin\phi}\alpha_x & \text{if $\vert\alpha_x\vert \ge \vert \alpha_y\vert$}\\
\frac{\sqrt{1+\epsilon^2}}{\sin\phi +\I \: \epsilon \cos\phi}\alpha_y & \text{else,}
\end{cases}
\end{align}\label{eq: polarizationconversion}
\end{subequations}
where for $c$ we picked the numerically most stable expression. Numerically one will run into problems using the above conversion if the emitted radiation is either linearly or circularly polarized. Therefore one should check beforehand if one of these conditions applies and use appropriate simplified conversion equations instead. 

\subsection{Polarization direction}

\begin{figure}[tbp]
\includegraphics[width=0.8\columnwidth]{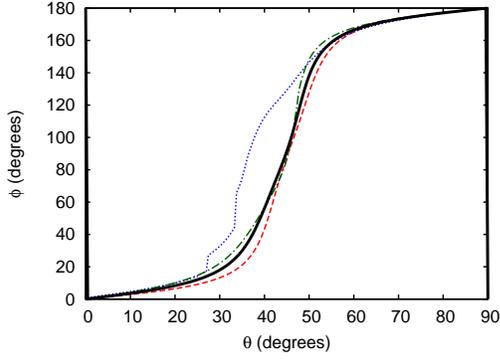}
\caption{(Color online) Main polarization direction $\phi$ for harmonic 49 generated by a Gaussian wave packet (black solid line) and generated by a three-cycle $\sin^2$-pulse (red dashed line), a five-cycle $\sin^2$-pulse (blue dotted line), and a fifteen-cycle trapezoidal pulse (green dot-dashed line).}
\label{fig: polarization1}
\end{figure}

The two-center interference minimum can also be observed in the polarization direction $\phi$ of the harmonics. Because the emission in the direction parallel to the laser polarization direction is strongly suppressed at the minimum, we expect a $\tfrac{\pi}{2}$-jump toward the minimum. The $\pi$-jump for the harmonic phase in the $x$-direction in \fig\ref{fig: harmonicphase} translates to a full $\pi$-rotation for $\phi$. This is exactly what is observed for a Gaussian wave packet and harmonics generated by different laser pulses in \fig\ref{fig: polarization1}. The polarization direction of the emitted radiation was averaged over one harmonic order using the total emitted intensities as weights.

\begin{figure}[tbp]
\includegraphics[width=0.8\columnwidth]{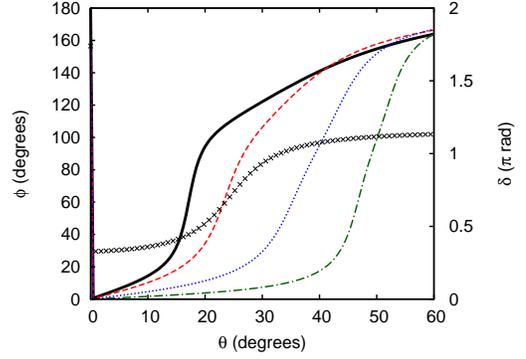}
\caption{(Color online) Main polarization direction $\phi$ for harmonic 31 (black solid line), harmonic 41 (red dashed line), harmonic 51 (blue dotted line), and harmonic 61 (green dot-dashed line). The black crosses are the phase difference $\delta$ for harmonic 31. A three-cycle $\sin^2$-pulse was used.}
\label{fig: polarization2}
\end{figure}

The phase jump in the $x$-direction at the two-center interference minimum becomes smoother for low harmonics \cite{Boutu08, Ciappina07, Chirila09}. In the following we investigate how the jump in the polarization direction depends on harmonic order. We plot the main polarization direction for different harmonics in \fig\ref{fig: polarization2}. 
Here we plot the polarization data points corresponding to the exact harmonics, \ie no averaging was done. The right-most curves for high harmonics show a simple polarization-direction jump around the two-center minimum. For the lowest harmonics (on the left-hand side) the behavior becomes more complicated. To explain this finding, we also plot the phase difference $\delta$ between $\alpha_y$ and $\alpha_x$ for harmonic 31 (black crosses). The intensity ratio $\tfrac{\vert \alpha_y\vert}{\vert \alpha_x \vert}$ equals 1 for $\theta=16\degree$ and $\theta=37\degree$, and it reaches a maximum of $\tfrac{\vert \alpha_y\vert}{\vert \alpha_x \vert}=2$ at $\theta=25\degree$. For a given ratio $\tfrac{\vert \alpha_y\vert}{\vert \alpha_x \vert}$, the main polarization direction is aligned more along the laboratory-frame ($x$ or $y$) direction with the higher amplitude if the phase difference $\delta$ between the laboratory-frame directions is far from 0 or $\pi$. We can observe this effect clearly in \fig\ref{fig: polarization2}: at $\theta$ equal to $20\degree$--$25\degree$ for harmonic 31, where $\delta$ is around $\tfrac{\pi}{2}$ and $\vert \alpha_y\vert$ is bigger than $\vert \alpha_x \vert$, the relatively slow increase in the polarization direction shows the tendency that the polarization is clamped toward the $y$-direction ($\phi=\tfrac{\pi}{2}$).        

\begin{figure}[tbp]
\includegraphics[width=0.8\columnwidth]{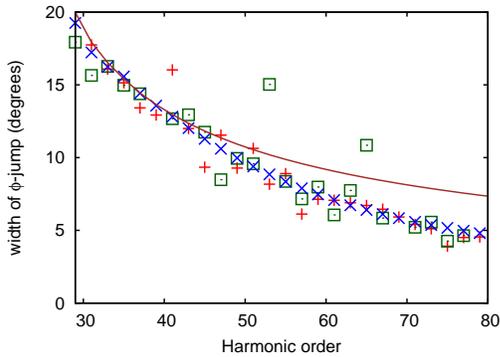}
\caption{(Color online) Width of the jump in $\phi$. Blue crosses are for the Gaussian wave packet, red plusses for the three-cycle pulse, and green squares for the fifteen-cycle pulse. The brown line shows the curve expected from purely geometric considerations.}
\label{fig: polarization3}
\end{figure}

We fit the jump observed in \fig\ref{fig: polarization1} with a smoothened step function to determine the location $\theta_{\mathrm{p}}$ and the width $\Delta \theta$ of the polarization-direction jump. When one plots the location $\theta_{\mathrm{p}}$ as a function of harmonic order, one obtains a result very similar to that shown in \fig\ref{fig: intensity2}. In \fig\ref{fig: polarization3} we plot the width $\Delta \theta$ as a function of harmonic order for a Gaussian wave packet and different laser pulses. 
Again, the results for the different laser pulses are scattered around the Gaussian-wave-packet result. The width of the jump for $\phi$ as a function of harmonic order does not depend on the laser pulse length. One observes that the width of the jump decreases with increasing harmonic order. An important part of this decrease is due to a purely geometric effect: with increasing harmonic order, the minimum moves to higher $\theta$ leading to a narrower interference pattern as a function of $\theta$. 

In the absence of Coulomb effects, \ie when $\alpha_x$ and $\alpha_y$ have the same phases apart from $\pi$-jumps, the polarization direction $\phi$ is given by $\tan\phi=\tfrac{\alpha_y}{\alpha_x}$. We assume that $\alpha_y$ is alignment independent over the range of the two-center minimum, as also indicated by numerical tests that we have performed, and that the alignment dependence of $\alpha_x$ comes purely from the two-center interference with the phase lag $k R \cos\theta$, \ie $\alpha_x=\alpha_x^{(0)} \cos\left(\tfrac{1}{2} k R \cos\theta \right)$ \cite{Ciappina07,Zhou08}. Then the variation $\Delta\phi$ of the polarization direction on varying $\theta$ in the vicinity of $\theta_{\mathrm{p}}$ is proportional to $\Delta\!\left(k R \cos\theta\right) \simeq \left(\pi/\cos\theta_{\mathrm{p}}\right)\Delta\!\left(\cos\theta\right)$. Thus the width $\Delta\theta$ should be such that $\Delta\!\left(\cos\theta\right)/\cos\theta_{\mathrm{p}}$ is independent of harmonic frequency, provided that $\alpha_y/\alpha_x^{(0)}$ is frequency independent. The brown line in \fig\ref{fig: polarization3} is obtained for $\Delta\theta$ if we set $\Delta\!\left(\cos\theta\right)/\cos\theta_{\mathrm{p}}$ arbitrarily equal to $0.185$ using $\Delta\!\left(\cos\theta\right)\simeq \cos \left(\theta_{\mathrm{p}}-\tfrac{\Delta \theta}{2}\right)- \cos \left(\theta_{\mathrm{p}}+\tfrac{\Delta \theta}{2}\right)$. A comparison between the brown line and the other curves shows that at the high end of the spectrum the decrease in $\Delta \theta$ cannot be explained any more exclusively by the geometric effect. Since $\theta_{\mathrm{p}}$ varies very slowly in this range, we expect only a slow variation in $\alpha_y/\alpha_x^{(0)}$. This suggests that decreasing Coulomb effects play a role, in accordance with \fig\ref{fig: intensity2}, where for harmonics 50--80 we observe a transition toward the curve predicted by the SFA dispersion relationship, also indicating decreasing Coulomb effects in this range.

\subsection{Ellipticity}

\begin{figure}[tbp]
\includegraphics[width=0.8\columnwidth]{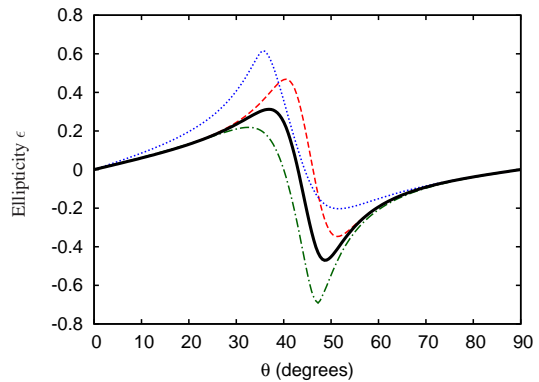}
\caption{(Color online) Ellipticity of the harmonic emission for harmonic 49 generated by a Gaussian wave packet (black solid line) and generated by a three-cycle $\sin^2$-pulse (red dashed line), a five-cycle $\sin^2$-pulse (blue dotted line), and a fifteen-cycle trapezoidal pulse (green dot-dashed line).}
\label{fig: ellipticity1}
\end{figure}

\begin{figure}[tbp]
\includegraphics[width=1.0\columnwidth]{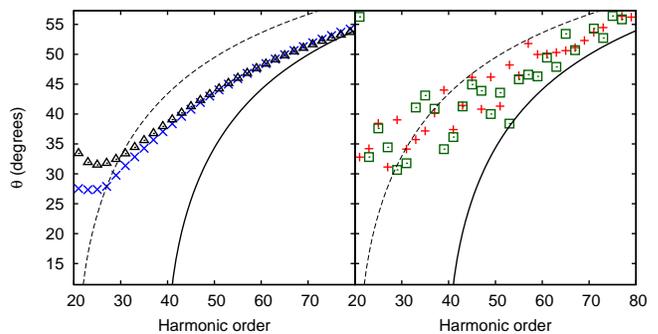}
\caption{(Color online) Alignment angles of zero ellipticity. (Left) Blue crosses are for a broad Gaussian wave packet starting far from the nucleus and black triangles for a narrow Gaussian wave packet starting close to the nucleus. (Right) Red plusses for the three-cycle pulse and green squares for the fifteen-cycle pulse.}
\label{fig: ellipticity2}
\end{figure}

In \fig\ref{fig: ellipticity1} we plot the ellipticity $\epsilon$ of the emitted radiation for harmonic 49 as a function of $\theta$. Again the polarization data was averaged over one harmonic order using the intensities as weights. The plot shows that both a Gaussian wave packet and different laser pulses give rise to both significant and varying elliptical emission. Nonzero ellipticity means that the harmonics in the $x$ and $y$-directions are emitted with different phases. Using the plane-wave approximation for the returning electron, one would not expect to see any ellipticity for a symmetric molecule \cite{Levesque07_2}. Since the ionization and propagation step are identical for the two components of the radiation, the ellipticity must come from the recombination step. Therefore, this result confirms that the Coulomb effects can lead to significant ellipticity. The ellipticity for parallel or perpendicular alignment is zero, because at these alignment angles, the perpendicular component of the emitted radiation vanishes. 
The ellipticity goes through zero at some intermediate alignment angle. If we plot the angle of zero ellipticity as a function of harmonic order, we arrive at \fig\ref{fig: ellipticity2}.
This plot shows that in the close vicinity of the two-center interference minimum, the ellipticity goes through zero. This is as expected, because at the location of the minimum, the $x$-component of the emitted radiation is very small. Because the $x$-component has opposite signs before and after the minimum, the ellipticity changes handedness through the minimum.

\begin{figure}[tbp]
\includegraphics[width=0.8\columnwidth]{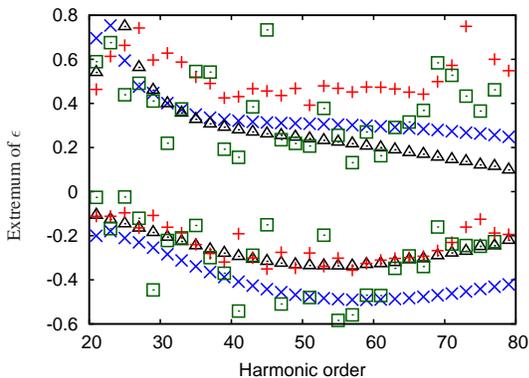}
\caption{(Color online) Extrema of ellipticity as a function of harmonic order. Blue crosses represent a broad Gaussian wave packet starting far from the nucleus, black triangles a narrow Gaussian wave packet starting close to the nucleus, red plusses a three-cycle pulse, and green squares a fifteen-cycle pulse.}
\label{fig: ellipticity3}
\end{figure}

\begin{figure}[tbp]
\includegraphics[width=0.8\columnwidth]{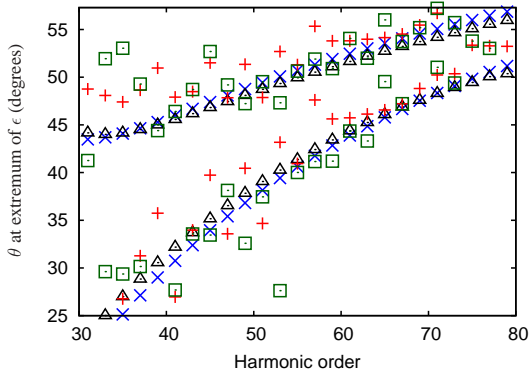}
\caption{(Color online) Alignment angles for extrema from \fig\ref{fig: ellipticity3}. Same wave packets and pulses as in \fig\ref{fig: ellipticity3}.}
\label{fig: ellipticity4}
\end{figure}

For each harmonic, we can also plot the extrema of the ellipticity that can be reached and the alignment angles at which those extrema are reached. The results are shown in \figs\ref{fig: ellipticity3} and \ref{fig: ellipticity4}, respectively.
We observe that for sufficiently high harmonic orders, the ellipticity extrema become smaller in absolute value and move closer to the two-center interference minimum with increasing harmonic order. Since the ellipticity is an indicator of non-plane-wave character, the decreasing ellipticity is another signature of decreasing Coulomb effects for higher harmonic orders, which correspond to higher return momenta.

\section{Conclusion}

When a Gaussian wave packet collides with an aligned H$_2^+$-molecule, the resulting harmonic spectrum has a structural minimum from the two-center interference between the two lobes of the orbital. The position of this minimum is reproduced using an effective plane-wave momentum that transitions from the $I_{\mathrm{p}}$-corrected $k(\omega)=\sqrt{2\omega}$ at low harmonics to the SFA-based $k(\omega)=\sqrt{2(\omega-I_{\mathrm{p}})}$ at high harmonics. A laser-induced HHG spectrum shows the same behavior if only a single electronic trajectory contributes per harmonic, as is the case for a typical experimental setup. This justifies using HHG for molecular imaging as the laser field has no significant effect on the amplitude of the recombination matrix element. When a single set of short and long electron trajectories contributes to the spectrum, the interference between the two trajectories causes a large but regular oscillation around the general trend. Introducing more and longer trajectories by using longer pulses has the effect of averaging out the oscillations to a smaller scattering around the Gaussian-wave-packet result and leads to a smoother interference minimum as a function of $\theta$.

Our results show that the effect of the Coulomb potential can lead to significant ellipticity of the emitted radiation. Around the interference minimum, the main polarization angle makes a $\pi$-jump and the ellipticity goes through zero. The Coulomb effects are less important at higher harmonics. Therefore we observe decreasing overall ellipticity and a relatively sharp jump in the polarization direction at the high end of the spectrum. 

In the wave-packet calculations, the Coulomb effects could be investigated in more detail by changing the strength of the potential for the evolution of the continuum part. This may be subject of future work. Finally, we mention that for randomly oriented molecules, the perpendicular harmonic components and thereby also the ellipticity vanish due to the cylindrical symmetry of the system around the laser polarization axis. The behavior of the phase is more complicated. Since, however, HHG is dominated in our case by the large orientation angles for geometrical reasons, we do not expect a clear signature of the phase jump for randomly oriented molecules.

\begin{acknowledgments}
The authors thank the Deutsche Forschungsgemeinschaft for funding the {\em Centre for Quantum Engineering and Space-Time Research} (QUEST). We acknowledge the support from the European Marie Curie Initial Training Network Grant No.\ CA-ITN-214962-FASTQUAST.
\end{acknowledgments}

\end{document}